\documentclass[aps,prl,floatfix,nofootinbib,twocolumn]{revtex4}

\usepackage{amssymb}
\usepackage{amsmath}
\usepackage{amsfonts}
\usepackage{graphicx}
\usepackage{color}
\usepackage{xspace}
\usepackage{ulem}
\usepackage{color}

\newcommand{\AddrAHEP}{
AHEP Group, Instituto de F\'{\i}sica Corpuscular -- C.S.I.C./Universitat de Val{\`e}ncia \\
c/ Catedr\'atico Jos\'e Beltr\'an, 2,
E- 46980 Paterna (Val{\`e}ncia) Spain
}

\newcommand{\AddrUCL}{
Department of Physics and Astronomy, University College London,\\
London WC1E 6BT, United Kingdom
}

\begin{document}


 \title{Is charged lepton flavour violation a high energy phenomenon?}

\author{Frank~F.~Deppisch}\email{f.deppisch@ucl.ac.uk}\affiliation{\AddrUCL}
\author{Nishita~Desai}\email{n.desai@ucl.ac.uk}\affiliation{\AddrUCL}
\author{Jos\'e~W.~F.~Valle}\email{valle@ific.uv.es}\affiliation{\AddrAHEP}


\begin{abstract}\noindent
  Searches for rare processes such as $\mu\to e\gamma$ put stringent
  limits on lepton flavour violation expected in many Beyond the
  Standard Model physics scenarios. This usually precludes the
  observation of flavour violation at high energy colliders such as
  the LHC. We here discuss a scenario where right-handed neutrinos are
  produced via a $Z'$ portal but which can only decay via small
  flavour violating couplings. Consequently, the process rate is
  unsuppressed by the small couplings and can be visible
  despite unobservably small $\mu\to e\gamma$ rates.
\end{abstract}

\maketitle

\section{Introduction}
\label{sec:intro}

Despite a long history of searches for lepton flavour violation (LFV)
in charged leptons no such signal beyond the Standard Model (SM) has
ever been observed, with the most impressive limit $Br(\mu\to e\gamma)
< 5.7\times 10^{-13}$ reported recently by the MEG
Collaboration~\cite{Adam:2013mnn}. The existence of neutrino
oscillations suggests that, at some level, LFV should also take place
in such processes. With only the light neutrinos of the SM present,
LFV is strongly suppressed by $\Delta m^2_\nu/m_W^2 \approx 10^{-50}$,
due to the Glashow–-Iliopoulos–-Maiani (GIM) mechanism. 
This results in LFV process rates far below
any experimental sensitivity which can be safely ignored. On the other
hand, many Beyond the Standard Model scenarios predict new sources for
charged lepton flavour violation associated, for example, with the
exchange of neutral heavy leptons or supersymmetric states. Since
these LFV processes can proceed even in the limit of strictly massless
neutrinos, their rates are unconstrained by the smallness of neutrino
masses~\cite{bernabeu:1987gr}.  However in many such scenarios the
non-observation of $\mu\to e\gamma$ in the MEG experiment places
stringent limits on expected LFV rates at high energies. As a
consequence, the observation of LFV at high energy colliders such as
the LHC is precluded due to either the mediating particles being too
heavy or their couplings to the charged leptons too small.

There are a few phenomenological solutions to this problem. For
example, the suppression with respect to the mass difference of the
LFV mediating particles differs depending on whether they are virtual
or produced as real resonances. In the latter case, the suppression is
only $\Delta M^2/(M \Gamma)$ instead of a GIM-like $\Delta
M^2/M^2$. This was demonstrated in the case of mediating heavy
neutrinos and a right-handed $W$ boson in Left-Right symmetrical
models in \cite{AguilarSaavedra:2012fu, Das:2012ii}. We here discuss
an alternative scenario where right-handed neutrinos are pair-produced
via a $Z'$ portal and can only decay via the flavour-dependent Yukawa
coupling to the light neutrinos \cite{delAguila:2007ua}. Consequently,
the process rate is unsuppressed by the small flavour couplings and
hence can be observable despite unobservably small $\mu\to e\gamma$
rates. Such a scenario can be realized in models with an extra U(1) or
extended gauge sector such as left-right symmetric models
\cite{malinsky:2005bi}.

\section{Neutrino Seesaw mechanism}
\label{sec:seesaw}

Within the standard type-I seesaw scenario with the mass matrix 
\begin{align}
\label{eq:seesaw}
		\begin{pmatrix}
				0   & m_D \\
				m_D & M_N
		\end{pmatrix},
\end{align}
for the left- and right-handed neutrino, the mixing between the
heavy $N$ and light $\nu$ states is completely fixed once the masses
are specified, $\theta \equiv m_D / m_N \approx \sqrt{m_\nu / m_N}$
\cite{minkowski:1977sc}. Eq.~\eqref{eq:seesaw} is expressed in terms
of one generation; for detailed multi-generation seesaw expansion
formulas for the mixing coefficients see Ref.~\cite{Schechter:1981cv}.
For the observed light neutrino mass scale $m_\nu \approx 0.1$~eV and
a TeV scale heavy neutrino, the mixing is negligibly small, $\theta
\approx 10^{-7}$ and, despite the breakdown of the GIM mechanism
\cite{Lee:1977tib}, the heavy neutrinos do not enhance low energy LFV
process rates sufficiently for detectability. This can dramatically
change in extended seesaw scenarios with TeV scale heavy
neutrinos. For example the inverse seesaw mechanism
\cite{mohapatra:1986bd} is described by the mass matrix
\begin{align}
\label{eq:invseesaw}
		\begin{pmatrix}
				0   & m_D & 0    \\
				m_D & 0   & m_N  \\
				0   & m_N & \mu
		\end{pmatrix},
\end{align}
including a sequential singlet state $S$ as third entry. With the
additional freedom introduced by the small lepton number violating $\mu$ parameter, light neutrinos can be accommodated for any
value of $\theta \equiv m_D / m_N$ \cite{gonzalezgarcia:1988rw}. In
essence, the magnitude of neutrino mass becomes decoupled from the
strength of lepton flavour violation \cite{bernabeu:1987gr}. 
Alternatively, the light-heavy mixing can also be enhanced within
the standard minimal seesaw sector by choosing specific flavour
textures in the mass matrix of the type-I seesaw, see for example
\cite{kersten:2007vk, He:2009ua, Ibarra:2010xw}.

For definiteness here we focus on LFV in the electron-muon sector
induced by the mixing between isodoublet and isosinglet neutrinos,
via the corresponding Yukawa couplings. As a result, the heavy
neutrinos couple to charged leptons via their small isodoublet
components $\theta^{e,\mu}$, which we treat as free
parameters. It is convenient to write these couplings in terms of
an overall mixing strength, $\theta \equiv \sqrt{\theta^e \theta^\mu}$
and the ratio of mixing strengths, $r_{e\mu} \equiv \theta^e /
\theta^\mu$. These parameters are unrestricted by the
smallness of neutrino masses; however they are constrained by weak universality
precision measurements to be $\theta^{e,\mu}
\lesssim 10^{-2}$ \cite{PhysRevD.86.010001}. 
We do not take into account possible constraints on $\theta$ from neutrinoless
double beta decay searches. Although highly stringent for a
heavy Majorana neutrino, they are avoided in the presence
of cancellations, such as in the quasi-Dirac neutrino case.

\section{$Z'$ models}
\label{sec:zprime}

Various physics scenarios beyond the Standard Model predict different
types of TeV-scale $Z'$ gauge bosons associated with an extra U(1)
that could arise, say, from unified SO(10) or E(6) extensions. An
introduction and extensive list of references can be found in
Ref.~\cite{Langacker:2008yv}. Electroweak precision measurements
restrict the mass and couplings of a $Z'$ boson. For example, lepton
universality at the $Z$ peak places lower limits on the $Z'$ boson
mass of the order $\mathcal{O}(1)$~TeV \cite{Polak:1991pc} depending
on hypercharge assignments. From the same data, the mixing angle
between $Z'$ and the SM $Z$ is constrained to be $\zeta_Z <
\mathcal{O}(10^{-4})$. For a discussion of direct limits on $Z'$
masses see \cite{PhysRevD.86.010001}. Recent limits from searches at
the LHC will be discussed in more detail below.

In the following we work in a simplified U(1)$'$ scenario with
only a $Z'$ and $N$ present beyond the SM. For the mechanism
described here to work, it is crucial that there are no other
particles present through which the heavy neutrino can decay
unsuppressed. For definiteness we assume two reference model cases:
the SO(10) derived U(1)$'$ coupling strength with the charge
assignments of the model described in \cite{malinsky:2005bi}, and a
leptophobic variant where the U(1)$'$ charges of SM leptons are set
to zero.

\section{Low Energy Lepton Flavour Violation}
\label{sec:lowlfv}

In the scenario considered here, the LFV branching ratio for the
process $\mu\to e\gamma$ can be expressed as \cite{Deppisch:2010fr}
\begin{align}
\label{eq:Bllgamma}
Br(\mu\to e\gamma)         &= 3.6 \times 10^{-3} G^2_\gamma\left(\frac{m_N^2}{m_W^2}\right) \times \theta_{}^4, \\
\text{with } G_\gamma &= -\frac{2 x^3 + 5 x^2 - x}{4 (1 - x)^3} -
\frac{3 x^3}{2 (1 - x)^4}\log(x), \nonumber
\end{align}
where the loop function $G_\gamma(x)$ is of order one with the limits
$G_\gamma \to 1/8$ for $m_N \to m_W$ and $G_\gamma \to 1/2$ for $m_N
\gg m_W$.
This prediction should be compared with the current experimental
limit \cite{Adam:2013mnn},
\begin{align}
\label{eq:Bexpllgamma}
	Br_\text{MEG}(\mu\to e\gamma) &< 5.7 \times 10^{-13} \text{ (90\% C.L.)},
\end{align}
from the MEG experiment which aims at a final sensitivity of
$Br(\mu\to e\gamma) \approx 10^{-13}$. The expression
\eqref{eq:Bllgamma} therefore results in a current upper limit on the
mixing parameter $\theta \lesssim 0.5\times 10^{-2}$ for $m_N =
1$~TeV. In contrast, the mixing strength $\theta\approx 10^{-7}$
expected in the standard high-scale type-I seesaw mechanism
Eq.~\eqref{eq:seesaw} would lead to an unobservable LFV rate with
$Br(\mu\to e\gamma) \approx 10^{-31}$.

If the photonic dipole operator responsible for $\mu\to e\gamma$ and
also contributing to $\mu\to eee$ and $\mu-e$ conversion in nuclei is
dominant, searches for the latter two processes do not provide
competitive bounds on the LFV scenario at the moment. Depending on the
breaking of the additional U(1)$'$ symmetry, non-decoupling effects
may appear which can boost the effective $Z'e\mu$ vertex contributing
to $\mu\to eee$ and $\mu-e$ conversion in nuclei
\cite{Tommasini:1995ii}.

\section{Heavy Neutrinos from the $Z'$ Portal}
\label{sec:resultsLHC}

%
\begin{figure}
\centering
\includegraphics[clip,width=0.42\textwidth]{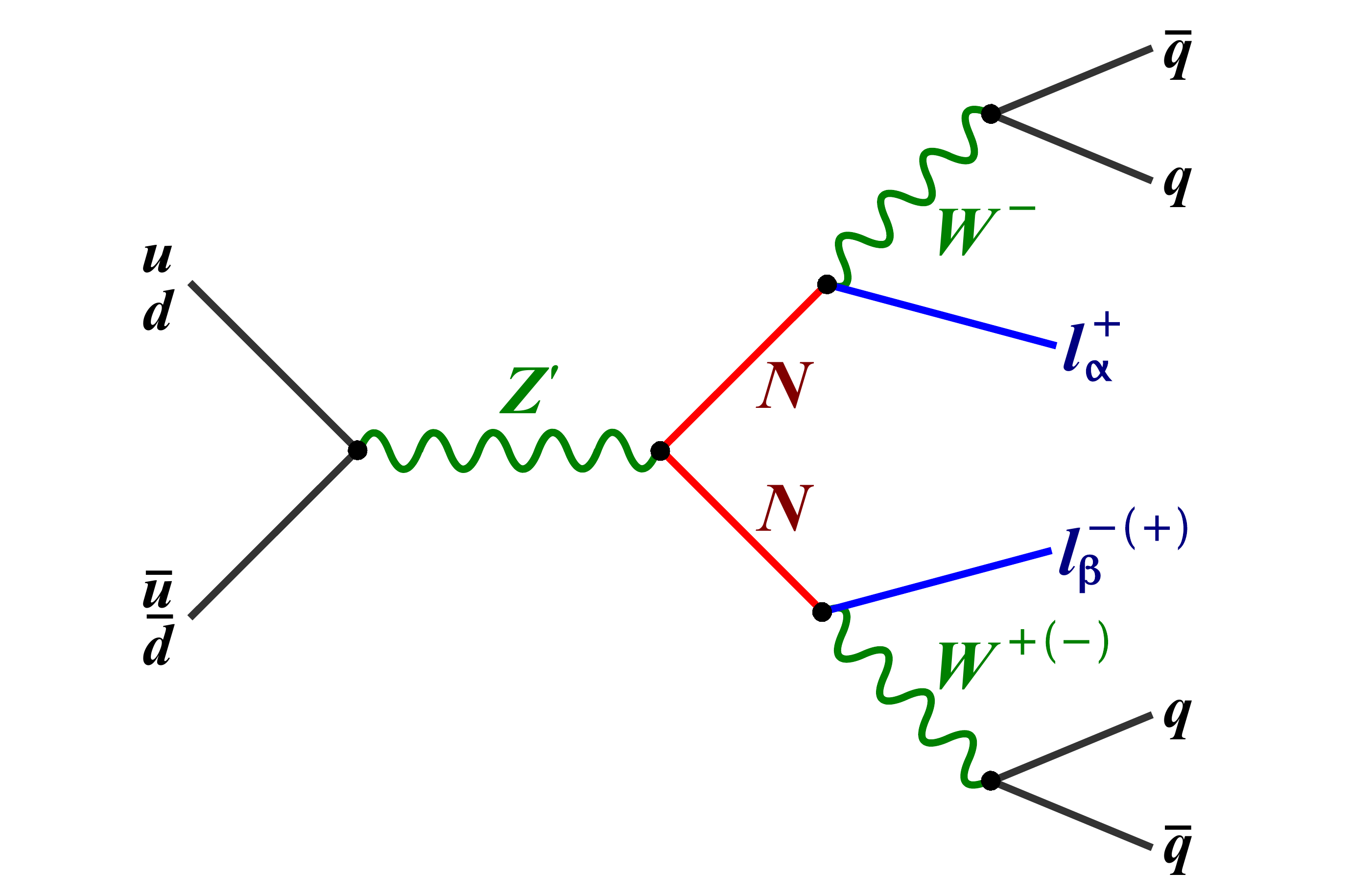}
\caption{Feynman diagram for heavy Majorana neutrino production through 
the $Z'$ portal at the LHC.}
\label{fig:diagram} 
\end{figure}
The process under consideration is depicted in
Figure~\ref{fig:diagram}. As shown, we will focus on the channel where
the heavy neutrinos decay into SM $W$ bosons which in turn decay
hadronically. The cross section of the production part $pp\to Z'$ can
be approximated by \cite{Leike:1998wr}
\begin{align}
\label{eq:ProdCrossApprox}
	\sigma(pp\to Z') & \approx 
		K\times C\times 
		\frac{4\pi^2}{3s}
		\frac{\Gamma_{Z'}}{m_{Z'}}
		\times \exp\left(-A \frac{m_{Z'}}{\sqrt{s}}\right) \nonumber\\
		&\times \left[ Br(Z'\to u\bar u)+ \frac{1}{2} Br(Z'\to d\bar d)\right],
\end{align}
with $C=600$, $A=32$ and the factor $K\approx 1.3$ describing higher
order QCD corrections. The target LHC beam energy is $\sqrt{s} =
14$~TeV. Here we focus on LFV at the LHC but not on lepton number
violation. The latter is usually considered as a smoking gun
signal of heavy Majorana neutrinos but realistic models with TeV scale
neutrinos such as inverse \cite{mohapatra:1986bd} and linear seesaw
\cite{malinsky:2005bi} scenarios usually lead to a quasi-Dirac nature
for the heavy neutrinos~\cite{Valle:1982yw}. It is strictly required
in case of large light-heavy mixing $\theta$ in order to ensure
adequately small neutrino masses $m_\nu\approx 0.1$~eV~\cite{kersten:2007vk}.
We therefore perform our calculations assuming a Dirac heavy neutrino
producing only opposite sign leptons. If it were a genuine Majorana
neutrino, inclusion of the same sign lepton signature would improve 
the discovery potential by taking advantage of the low background
expected for same sign lepton signatures. From this point of view the
results obtained here are conservative.

\begin{figure}
\centering
\includegraphics[clip,width=0.42\textwidth]{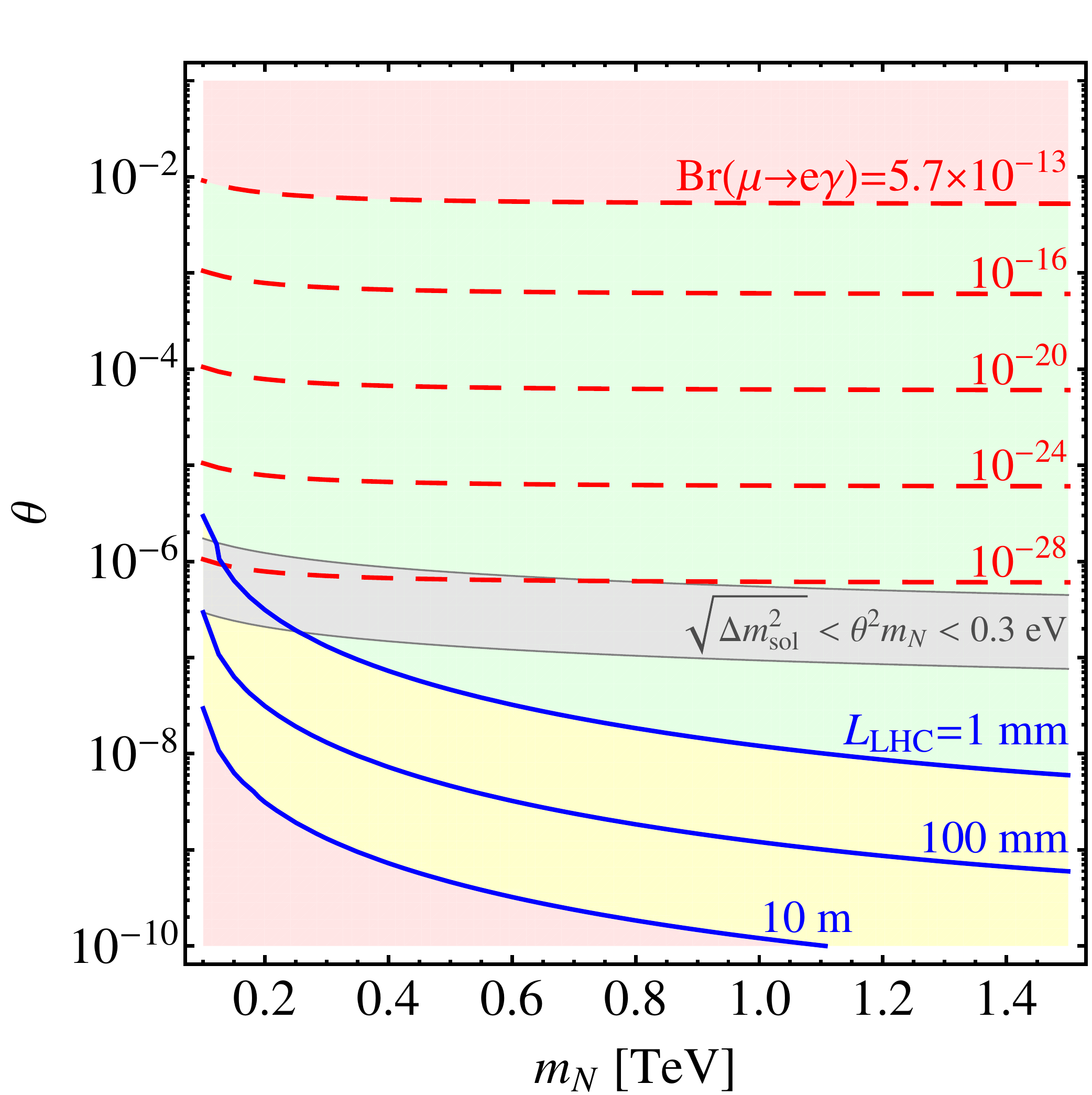}
\caption{Average decay length of a heavy neutrino $N$ produced in $Z' \to NN$ with $m_{Z'}=3$~TeV as a function of its mass $m_N$ and the light-heavy mixing $\theta$ (solid blue contours). The dashed red contours denote constant values for $Br(\mu\to e\gamma)$ whereas the grey shaded band corresponds to parameter values which produce light neutrino mass scales $m_\nu = \theta^2 m_N$ between $\sqrt{\Delta m^2_\text{sol}}$ and 0.3~eV within the canonical type-I seesaw mechanism.}
\label{fig:decaylength} 
\end{figure}
The total cross section of the LFV signal process $pp \to Z' \to N N
\to e^\pm \mu^\mp + 4\text{j}$ is then given by
\begin{align}
\label{eq:totalCS}
	\sigma_{e\mu} &= \sigma(pp\to Z') \times Br(Z' \to N N) \times Br(N \to e^\pm W^\mp) \nonumber\\
	&\times Br(N \to \mu^\mp W^\pm) \times Br^2(W^\pm \to 2 \text{j}).
\end{align}
The neutrino $N$ can decay via the channels $\ell^\pm W^\mp$,
$\nu_\ell Z$ and $\nu_\ell h$, all of which are suppressed by the
small mixing parameters $\theta^\ell$, $\ell=e,\mu$.  In the presence
of multiple heavy neutrinos with small mass differences we neglect the
decays involving either real or virtual $Z'$, $N_i \to N_j Z'$.
The branching ratio of the above channels into a given lepton
flavour is independent of the overall mixing strength $\theta$.

As long as the total decay width $\Gamma_N$ is large enough so that the 
heavy neutrino decays within the detector, the LHC LFV process rate is 
unsuppressed by the overall mixing strength $\theta$. 
The decay 
length of the heavy neutrino (in the rest frame of a 3~TeV $Z'$) is 
shown in Figure~\ref{fig:decaylength} as a function of $m_N$ and the 
light-heavy mixing $\theta$, in comparison with $Br(\mu\to e\gamma)$. 
For $\theta \gtrsim 10^{-7}$ and $m_N \gtrsim 0.3$~TeV, the neutrino 
decays promptly with a decay length $L < 1$~mm, and the LHC LFV process 
considered here is independent of and completely unsuppressed by $\theta$. 
The inclusion of the $Z'$ boost in the detector frame does not significantly 
alter this conclusion, but in general leads to a slight broadening of the 
yellow region. For lengths between 1~mm - 10~m, the $N$ decay may still 
be observable with potentially spectacular signatures such as displaced 
vertices or in-detector decays. Figure~\ref{fig:decaylength} also indicates 
the parameter area corresponding to the observed neutrino mass scale 
$m_\nu = \theta^2 m_N$ in the standard type-I seesaw mechanism, clearly 
showing that this regime cannot be probed by low energy searches but 
potentially by the LHC process considered here.

The total cross
section \eqref{eq:totalCS} only depends on the ratio $r_{e\mu}$ of the
flavour couplings, $\sigma_{e\mu} \propto r^2_{e\mu} /
(r^2_{e\mu}+1)^2$, and is maximal for $r_{e\mu}=1$.  This is very much
in contrast to $Br(\mu\to e\gamma)$ in Eq.~\eqref{eq:Bllgamma} which
is heavily suppressed by a small value of $\theta$, though is
independent of the ratio $r_{e\mu}$.

\begin{figure}
\centering
\includegraphics[clip,width=0.40\textwidth]{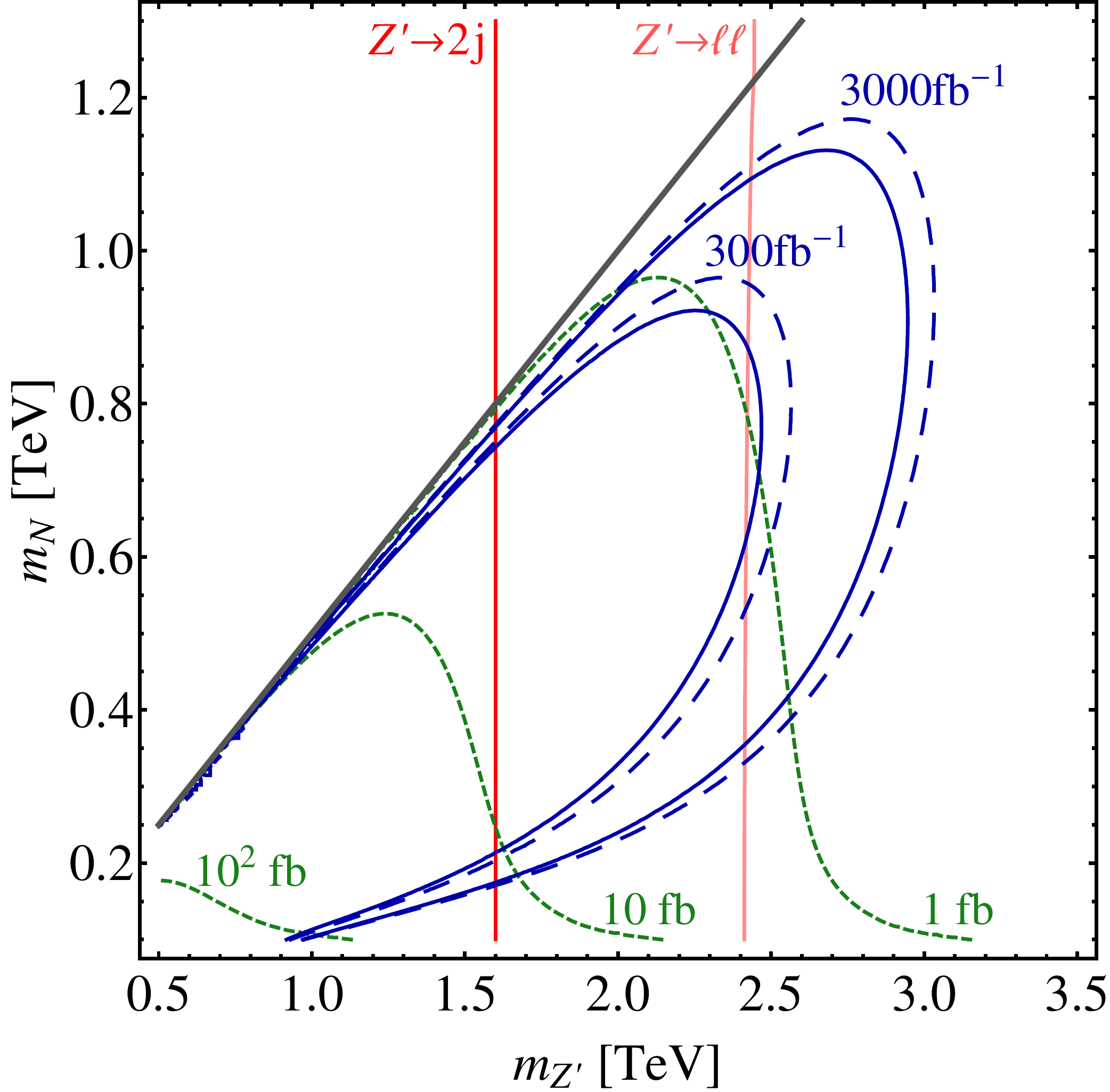}
\caption{Cross section $\sigma(pp \to Z' \to N N \to e^\pm \mu^\mp +
  4\text{j})$ at the LHC with 14~TeV as a function of $m_{Z'}$ and
  $m_N$ for maximal LFV (dotted contours). The solid and long dashed
  contours give the required luminosity at the LHC for a $5\sigma$ 
	discovery, in the case of SO(10) and leptophobic
  charges, respectively. The vertical lines denote the upper limit on
  $m_{Z'}$ from existing LHC searches in dijet and dilepton channels.}
\label{fig:mZp_mN} 
\end{figure}
In order to examine the viability of observing the signal at the 14~TeV 
run of the LHC we perform a simulation of $pp \to Z' \to NN \to \ell_1^+ 
W^- \ell_2^- W^+$ using \textsc{Pythia}~8 \cite{Sjostrand:2007gs} with both $W$ 
bosons decaying into quarks producing a $2 \ell + 4\mathrm{j}$ final state. 
Possible SM backgrounds arise from the channels $(t\bar t,~Z,~tW,~WW,~WZ,~ZZ) 
+ n\text{j}$ which we simulate using \textsc{Madgraph}~5 \cite{Alwall:2011uj}. We 
include parton showering and hadronization for both signal and background
using \textsc{Pythia}~8.
We apply the following selection criteria: (i) An event must have four
jets with a transverse momentum of at least 40~GeV each and (ii) two
opposite-sign leptons with transverse momenta $p_T > 120$~GeV; (iii)
since there is no source of missing transverse energy (MET) in the
signal, we require $\text{MET} < 30$~GeV and (iv) a large dilepton
invariant mass $M_{\ell \ell} > 400$~GeV further reduces the $t\bar t$
background and reduces the $Z+n\text{j}$ and $V V+n\text{j}$ to
negligible amounts. 
In addition, the heavy neutrino mass could
be determined through a peak in the invariant mass $m_{\ell jj}$,
although the sharpness of such a peak is likely to be reduced due to
the combinatorics of identifying the correct final particles.

Figure~\ref{fig:mZp_mN} shows the cross section of the process $pp \to
Z' \to N N \to e^\pm \mu^\mp + 4 \text{j}$ at the LHC with 14~TeV as a 
function of $m_{Z'}$ and $m_N$ for maximal LFV, $r_{e\mu} = 1$. In
addition, it provides an estimate of the required luminosity at the LHC
to observe a $5\sigma$ LFV signal over background significance, 
as derived using the simulation procedure
described above. In addition to the case with SO(10) derived U(1)$'$
charges, it also shows the expected significance for a leptophobic
$Z'$ with the lepton doublet and charged lepton singlet charges put to
zero. This increases the signal cross section by about 25\% due to the
increased $Z'$ decay branching ratio into heavy neutrinos. We find
that LFV can potentially be discovered for heavy neutrinos and $Z'$
with masses $m_N \lesssim 0.9$~TeV and $m_{Z'} \lesssim 2.5$~TeV,
respectively. In the case of three degenerate neutrinos with identical
$r_{e\mu}=1$, this reach would increase to $m_N \lesssim 1.1$~TeV and
$m_{Z'} \lesssim 3.0$~TeV.

In determining the LHC potential to discover LFV through the process 
considered here, we must take into account existing $Z'$ LHC searches. 
The vertical lines in Figure~\ref{fig:mZp_mN} indicate the upper limits 
on $m_{Z'}$ from the LHC 8~TeV run in the dijet channel $pp \to Z' \to 
2\text{j}$ \cite{ATLAS:2012qjz} (assuming SM charges and couplings) and 
the dilepton channel $pp \to Z' \to \ell^+\ell^-$, $\ell = e,\mu$ 
\cite{ATLAS:2013jma} (assuming SO(10) derived couplings and charges). 
The corresponding limit from dilepton searches reported by CMS 
\cite{CMS-PAS-EXO-12-061} is slightly stronger with $m_{Z'} \gtrsim 2.6$~TeV 
but difficult to consistently apply in our case as it is quoted only in 
terms of the cross section ratio to the SM $Z$ production.
The parameter space of the scenario with SO(10) derived charges is strongly 
constrained by dilepton searches. On the other hand, the leptophobic scenario, 
only limited by the dijet searches, still allows a large parameter space where 
a strong LFV signature could be observed.

\begin{figure}
\centering
\includegraphics[clip,width=0.40\textwidth]{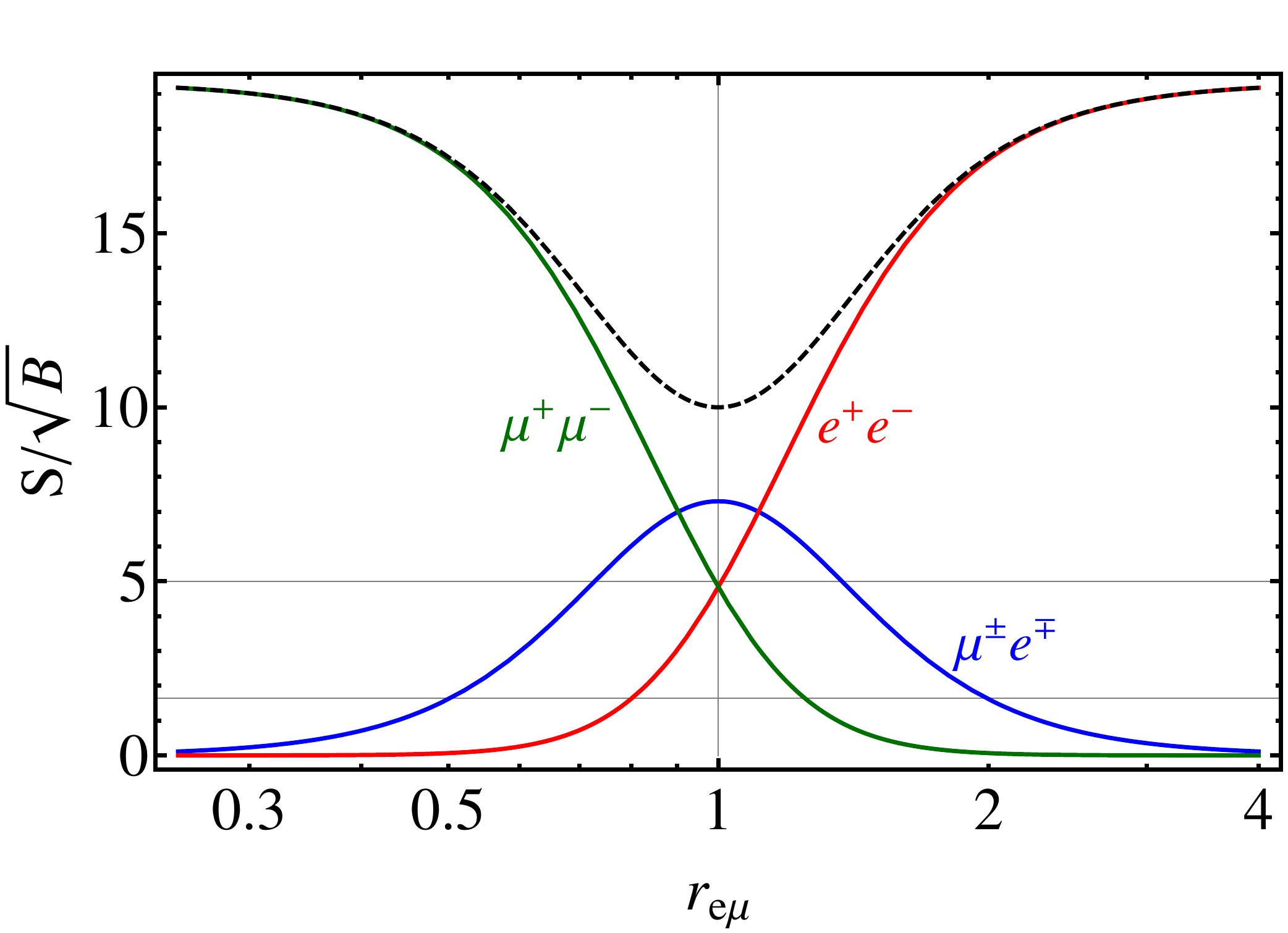}
\caption{Signal over background significance of $\sigma(pp \to Z' \to N N \to \ell\ell+ 4\text{j})$
  ($\ell\ell = \mu^\pm e^\mp, e^+ e^-, \mu^+ \mu^-$) at the LHC with 14~TeV and $\mathcal{L} = 300\text{ fb}^{-1}$
  as a function of $r_{e\mu}$. The masses are $(m_{Z'},m_N) = (2.4,0.75)$~TeV. The dashed black curve 
	gives the significances added in quadrature and the horizontal lines denote $5\sigma$ and 90\% significance thresholds.}
\label{fig:remu} 
\end{figure}
The effect of the coupling ratio $r_{e\mu}$ on all three flavour
channels $\mu^\pm e^\mp$, $e^+ e^-$ and $\mu^+ \mu^-$ is shown in
Figure~\ref{fig:remu} where the signal significances as well as their
sum in quadrature are plotted. Strongly non-universal couplings,
i.e. with the neutrino coupling dominantly to either $e$ or $\mu$, result in
the largest overall significance as the flavour content of the
background is $N(\mu^\pm e^\mp) : N(e^+ e^-) : N(\mu^+ \mu^-) \approx
2:1:1$. In contrast, the unambiguous discovery of LFV requires
approximately universal couplings, $r_{e\mu} \approx 1$.

\section{Conclusions}
\label{sec:conclusion}

The seesaw mechanism and its low-scale variants provide a well
motivated scenario for neutrino mass generation in many new physics
models. The experimental non-observation of low energy lepton flavour
violating processes puts stringent constraints on the scale and the
flavour structure of such models. This usually means that the
discovery of related LFV processes or heavy resonances at the LHC is
already ruled out. Here we discussed a scenario with negligible lepton
flavour violating rates in low energy rare process, while testable at
the high energies accessible at the LHC. 
The scenario described here illustrates a general mechanism, namely, 
(i) a LFV messenger particle is produced through a portal via an 
unsuppressed coupling but (ii) can only decay via small lepton flavour 
violating couplings. Such a scenario would provide an alternative solution 
to the general flavour problem in Beyond-the-Standard Model physics which 
is testable at high energy colliders, despite tiny LFV couplings and 
consequently unobservably small low energy LFV rates.

\section{Acknowledgments}
\begin{acknowledgments}
The work of F.F.D. and N.D. was supported
partly by the London Centre for Terauniverse
Studies (LCTS), using funding from
the European Research Council via the Advanced
Investigator Grant 267352. F.F.D.
gratefully acknowledges support from a IPPP associateship.
The work of J.W.F.V. was supported by the Spanish MINECO under Grants No.
FPA2011-22975 and MULTIDARK No. CSD2009-00064 (Consolider-Ingenio 2010
Program), by Prometeo/2009/091 (Generalitat Valenciana), and by
the EU ITN UNILHC PITN-GA-2009-237920. F.F.D. and N.D. would like to thank
Robert Thorne for useful discussions.
\end{acknowledgments}


\end{document}